\documentclass[%
 reprint,
 amsmath,amssymb,
 aps,
]{revtex4-2}

\usepackage{graphicx}
\usepackage{dcolumn}
\usepackage{bm}

\begin{document}

\preprint{APS/123-QED}

\title{Anomalous dependence of sensitivity on observation time caused by memory retention in the time crystal}

\author{Timofey T. Sergeev}
\affiliation{Moscow Institute of Physics and Technology, 141700, 9 Institutskiy pereulok, Moscow, Russia}
\affiliation{Dukhov Research Institute of Automatics (VNIIA), 127055, 22 Sushchevskaya, Moscow, Russia}
\affiliation{Institute for Theoretical and Applied Electromagnetics, 125412, 13 Izhorskaya, Moscow, Russia}
\affiliation{Institute of Spectroscopy Russian Academy of Sciences, 108840, 5 Fizicheskaya, Troitsk, Moscow, Russia}
\author{Alexander A. Zyablovsky}
\email{zyablovskiy@mail.ru}
\affiliation{Moscow Institute of Physics and Technology, 141700, 9 Institutskiy pereulok, Moscow, Russia}
\affiliation{Dukhov Research Institute of Automatics (VNIIA), 127055, 22 Sushchevskaya, Moscow, Russia}
\affiliation{Institute for Theoretical and Applied Electromagnetics, 125412, 13 Izhorskaya, Moscow, Russia}
\author{Evgeny S. Andrianov}
\affiliation{Moscow Institute of Physics and Technology, 141700, 9 Institutskiy pereulok, Moscow, Russia}
\affiliation{Dukhov Research Institute of Automatics (VNIIA), 127055, 22 Sushchevskaya, Moscow, Russia}
\affiliation{Institute for Theoretical and Applied Electromagnetics, 125412, 13 Izhorskaya, Moscow, Russia}

\date{\today}

\begin{abstract}
  In this work, we consider a composite atom-cavity system interacting with a ring resonator. In such a structure, time crystal regime can be observed. We show that a quadratic observation time dependence of the system’s sensitivity to perturbations takes place in the time crystal regime and also in the transition area to the normal state. This dependence is due to ability of the system to retain the memory of the atom’s initial state. Outside these areas, the system is not able to retain the memory of the atom’s initial state and the sensitivity scales linearly on the observation time. Our results open up a new way for implementation of discrete time crystals in sensing and metrology.
\end{abstract}

\maketitle

\section{Introduction}
Recently, systems with spontaneously broken time translation symmetry and time crystals have attracted increasing interest. Originally, the concept of time crystals was proposed by F. Wilczek \cite{1}. This concept based on the phenomenon of spontaneous breaking of time translation symmetry. Despite being criticized initially \cite{2,3,4}, the concept was extended to periodically driven Hamiltonian systems \cite{5,6,7,8}. The external periodic force generates a discrete time translation symmetry $H(t)=H(t+T)$ with respect to time period of external driving $T$ \cite{7,8}. Spontaneous breaking of time translation symmetry consists in the fact that there is a quantity in the system that changes with a multiple period of external driving $T'=mT, m=2,3,...$ \cite{7,8}. There are a number of implementations where is possible to observe discrete time crystals, for example, driven atom-cavity structures \cite{9}, systems with trapped ions \cite{10}, NV-centers \cite{11}, NMR spin clusters \cite{13,14}, quantum dots and electron-nuclear spin structures \cite{34, 36}, single-molecule magnet arrays \cite{33}, BCS superconductors \cite{35, 38}, polariton condensates \cite{39, 37} and open optomechanical systems \cite{12}.

Discrete time crystals and structures with spontaneously broken symmetries in general have intriguing and promising applications in quantum computation \cite{21,22} and in quantum metrology and sensing \cite{16,17,18,19,31,20,32,27,28,29,30}. For example, time crystal-based quantum devices also demonstrate sensitivity scaling in particle number and time that exceeds the standard quantum and Heisenberg limits \cite{20,32,27,28,29,30}. This fact opens up a new possible way of using time crystals in sensors. 

In this work, we consider a system consisting of a rotating ring resonator coupled with single-mode cavity, in which an active atom is placed. In such a scheme the spontaneous breaking of time translation symmetry can be observed \cite{15}. We demonstrate that the influence of perturbations of system's parameters, i.e. the sensitivity, depends on an observation time as $T^2$ in the time crystal regime and the transition area between the time crystal and the normal state. On the other hand, in the normal state, the sensitivity demonstrates linear dependence on the observation time. This effect is explained by the fact that in the time crystal regime, the system is able to retain the memory of its initial atom's state, which leads to a quadratic dependence of the sensitivity to perturbation magnitude on the observation time. Outside the time crystal regime, the system is not able to retain the memory of the initial state, so the dependence on the observation time becomes linear. This result can be useful for enhancing sensitivity of structures with rotational-broken symmetry, e.g., optical gyroscopes.

\section{Model}

We consider a system consisting of an active two-level atom placed into the single-mode cavity coupled with a ring resonator of length $L= 2 \pi R$. We consider the transition frequency of the atom and the frequency of the single-mode cavity both equal to $\omega_0$. In what follows, we consider $N$ modes whose frequencies lie close to ${\omega _0}$ and assume that ${\omega _0}/\delta \omega  = {j_0}$ is an integer number. In the absence of perturbations, the ring resonator modes are divided into "clockwise" and "counterclockwise" modes, whose frequencies are given as $\omega _j^ \pm  = j\,\delta \omega $, where $\delta \omega  = 2\pi c/L$ is a step between the modes’ frequencies.  When the system under consideration rotates with angular frequency ${\Omega _{rot}}$, the frequencies of "clockwise" and "counterclockwise" modes take the form $\omega _j^ \pm  = j\,\delta {\omega ^ \pm } \approx j\,\delta \omega \left( {1 \pm \varepsilon } \right)$, where $\delta {\omega ^ \pm } = \frac{{2\pi \left( {c \pm {\Omega _{rot}} \cdot R} \right)}}{L} \approx \delta \omega \left( {1 \pm \frac{{{\Omega _{rot}} \cdot R}}{c}} \right)$ are steps between the modes’ frequencies for "clockwise" and "counterclockwise" and $\varepsilon  = {\Omega _{rot}} \cdot R/c <  < 1$. The shift in the mode frequencies caused by the rotation of the system can be considered as a perturbation. Such types of systems can be realized, for example, in optical resonator gyroscopes \cite{41, 42} or in photonic structures with coupled qubits \cite{40}.

We investigate the dynamics of the system for different values $\varepsilon $ and for different values of the coupling strength between the atom and the single-mode resonator. We use the following Hamiltonian for the analysis \cite{23}:

\begin{equation}
\begin{array}{l}
\hat H = {\omega _0}\hat \sigma^\dag {{\hat \sigma}} + {\omega _0}\hat a^\dag {{\hat a}} + \Omega (\hat a {{\hat \sigma^{\dag} }} + \hat a^{\dag} {{\hat \sigma}}) + \\
\sum\limits_{j = j_0-N/2}^{j_0+N/2} {{\omega_j^{+}}\hat \alpha_j^\dag {{\hat \alpha}_j}} + \sum\limits_{j = j_0-N/2}^{j_0+N/2} {{g_j}(\hat a {{\hat \alpha}_j^{\dag}} + \hat a^{\dag} {{\hat \alpha}_j})}  + \\
\sum\limits_{j = j_0-N/2}^{j_0+N/2} {{\omega_j^{-}}\hat \beta_j^\dag {{\hat \beta}_j}} +\sum\limits_{j = j_0-N/2}^{j_0+N/2} {{g_j}(\hat a {{\hat \beta}_j^{\dag}} + \hat a^{\dag} {{\hat \beta}_j})}
\end{array}
\label{eq:1}
\end{equation}
where $\hat \sigma$ and $\hat\sigma^{\dag}$ are annihilation and creation operators of two-level atom that obey the fermionic commutation relation $\{\hat\sigma,\hat\sigma^{\dag}\} = 1$. $\hat a$ and $\hat a^{\dag}$ are annihilation and creation operators of the single-mode cavity. $\hat \alpha_j$, $\hat \alpha_j^{\dag}$ and $\hat \beta_j$, $\hat \beta_j^{\dag}$ are annihilation and creation operators of "clockwise" and "counterclockwise" modes of the ring resonator, respectively. The operators of the single-mode cavity and the ring resonator modes obey the bosonic commutation relations $[\hat a, \hat a^{\dag}] = 1$, $[\hat \alpha_i, \hat \alpha_j^{\dag}] = \delta_{ij}$ and $[\hat \beta_i, \hat \beta_j^{\dag}] = \delta_{ij}$. $\Omega$ is the coupling strength between the single-mode cavity and the atom. We consider both groups of modes to interact with the single-mode cavity with the same coupling strength, i.e. $g_j = g$ for all $j$.

The system dynamics are determined by the time-dependent Schr\"{o}dinger equation for a wave function $\vert\Psi (t)\rangle$. We assume that there is only one excitation quantum that can be in the atom, in the single-mode cavity, or in one of the ring resonator "clockwise" and "counterclockwise" modes. We look for the wave function in the following form \cite{23}:

\begin{equation}
\begin{array}{l}
\vert \Psi(t)\rangle = C_{\sigma}(t)\vert e,0,0,0\rangle + C_{a}(t)\vert g,1,0,0\rangle + \\
\sum\limits_{j = j_0-N/2}^{j_0+N/2} {C_{j}^{+}(t)\vert g,0,1_{j}^{+},0\rangle} + \sum\limits_{j = j_0-N/2}^{j_0+N/2} {C_{j}^{-}(t)\vert g,0,0,1_{j}^{-}\rangle} 
\end{array}
\label{eq:2}
\end{equation}
where $\vert e,0,0,0 \rangle$, $\vert g,1,0,0 \rangle$, $\vert g,0,1_{j}^{+},0\rangle$ and $\vert g,0,0,1_{j}^{-}\rangle$ are the states, in which the excitation quantum is in the atom, in the single-mode cavity, in one of the ring resonator "clockwise" and "counterclockwise" modes, respectively. $C_{\sigma}(t)$, $C_{a}(t)$, $C_{j}^{+}(t)$ and $C_{j}^{-}(t)$ are the amplitudes of probability of finding the excitation quantum in the atom, in the single-mode cavity, or in one of the ring resonator "clockwise" and "counterclockwise" modes, respectively.

We can obtain the closed system of equations for the probability amplitudes after substituting the wave function~(\ref{eq:2}) into the Schr\"{o}dinger equation with Hamiltonian~(\ref{eq:1}):

\begin{equation}
\frac{{d{C_{\sigma}}}}{{dt}} =  - i{\omega _0}{C_{\sigma}} - i\,\Omega {C_{a}} 
\label{eq:3}
\end{equation}

\begin{equation}
\frac{{d{C_a}}}{{dt}} =  - i{\omega _0}{C_a} - i\,\Omega {C_{\sigma}}-i\sum\limits_{j = j_0-N/2}^{j_0+N/2} {g{C_j^{+}}}-i\sum\limits_{j = j_0-N/2}^{j_0+N/2} {g{C_j^{-}}}
\label{eq:4}
\end{equation}

\begin{equation}
\frac{{d{C_j^{+}}}}{{dt}} =  - i{\omega_j^{+}}{C_j^{+}} - ig{C_a}
\label{eq:5}
\end{equation}

\begin{equation}
\frac{{d{C_j^{-}}}}{{dt}} =  - i{\omega_j^{-}}{C_j^{-}} - ig{C_a}
\label{eq:6}
\end{equation}
We also assume that at the initial moment of time, the excitation quantum is in the atom, i.e. $C_{\sigma}(0)=1$ and $C_a (0)=C_j^{+}(0)=C_j^{-}(0)=0, j=j_0-N/2,...,j_0+N/2$. In the following evolution the atom emits a photon into the ring resonator. The electromagnetic wave makes a complete bypass of the ring resonator in time $T_R = L/c = 2\pi/{\delta\omega}$.

Note that the Eqns.~(\ref{eq:3})-(\ref{eq:6}) are identical in form to the equations for the system with a large number of atoms (see Appendix A). That is, the obtained results are applicable to a macroscopic system with a large number of atoms.

\section{Sensitivity of system dynamics to perturbations}

\begin{figure*}[htbp]
\centering\includegraphics[width=1.0\linewidth]{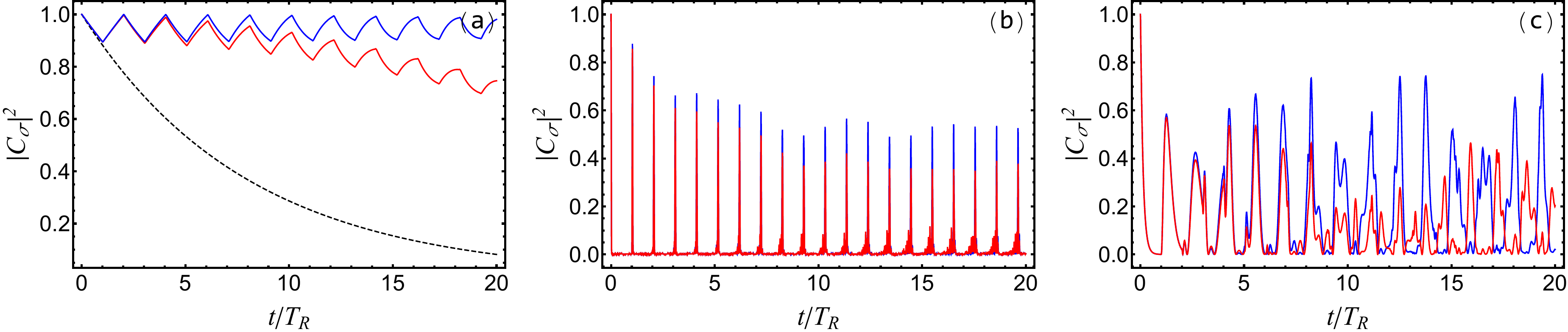}
\caption{Dependence of the atom's probability density $\vert C_{\sigma}(t)\vert^{2}$ on time in the case of zero perturbations $\varepsilon = 0$ (blue solid line) and in the case of non-zero perturbation $\varepsilon = 10^{-4}$ (red solid line) in the time crystal regime $\Omega = 0.25 \cdot \Omega_{TC}$ (a), outside the time crystal regime $\Omega = 8 \cdot \Omega_{TC}$ (b) and in the transition area $\Omega = 2 \cdot \Omega_{TC}$ (c). Black dashed line in (a) corresponds to function $f(t)=e^{-2\gamma't}$, where $\gamma'$ is the effective dissipation rate of the atom's probability amplitude. Here $T_R$ is the time of one bypass of the ring resonator. We consider $N=50$, $\delta\omega = 4\cdot 10^{-3} \omega_0$, $g=6\cdot 10^{-3}\omega_0$.}
\label{fig1}
\end{figure*}

The system under consideration is in the regime of time crystal when the coupling strength is less than the critical coupling strength \cite{15}. The critical coupling strength of the time crystal regime can be calculated in a similar way as presented in \cite{15}. The time crystal regime begins to disrupt when the $\gamma' T_{R} \thicksim 1$, where $\gamma'$ is effective dissipation rate of the atom's probability amplitude in the time crystal regime and $T_R$ is the time of one bypass of the ring resonator. $\gamma' = \Omega^{2} / \gamma$, where $\Omega << \gamma$ and $\gamma = \frac{2\pi g^2}{\delta\omega}$ is the atom's probability amplitude dissipation rate that is calculated in Born-Markovian approximation \cite{24, 25} with the elimination of the ring resonators' degrees of freedom \cite{15, 24, 25, 26}. Note that this approximation describes the dynamics only at times before the time of the first bypass (see black dashed line in Figure~\ref{fig1}(a), $t<T_R$). Thus, the condition $\gamma' T_{R} \thicksim 1$ can be written as $\Omega \thicksim g$. The value of the coupling strength $\Omega = \Omega_{TC} = g$ is the critical coupling strength of the time crystal regime.

The value of the effective dissipation rate $\gamma'$ gives an estimation for the effective time of relaxation $T_{eff} \sim 1/{\gamma'}$. However, in the time crystal regime the atom's probability amplitude does not dissipate and oscillates near initial state in any observable time interval (see Figure~\ref{fig1} (a)). Such oscillations with a period of $2T_R$ are persistent even at times much greater than the time of one bypass (see Figure~\ref{fig1a} in Appendix B). In contrast, outside the time crystal regime the atom's probability amplitude completely dissipates in the time interval that is less then $T_R$ (Figure~\ref{fig1} (b)). This effect indicates the ability of the system to retain the memory of the atom's state in the time crystal regime and its absence outside the time crystal regime.

The presence of perturbations leads to change in the system's dynamics [Figure~\ref{fig1}]. To characterize the influence of the perturbation, we calculate the relative difference between the time-averaged probability densities of the atom for zero and nonzero perturbation magnitudes. We determine a sensitivity of the system to the magnitude of perturbation as

\begin{equation}
S(T,\varepsilon,\Omega)=1- \int_0^T \vert C_{\sigma,\varepsilon \neq 0} \vert^{2} {dt} / \int_0^T \vert C_{\sigma,\varepsilon = 0} \vert^{2} {dt}
\label{eq:7}
\end{equation}
where $T$ is an observation time. 

\begin{figure*}[htbp]
\centering\includegraphics[width=0.8\linewidth]{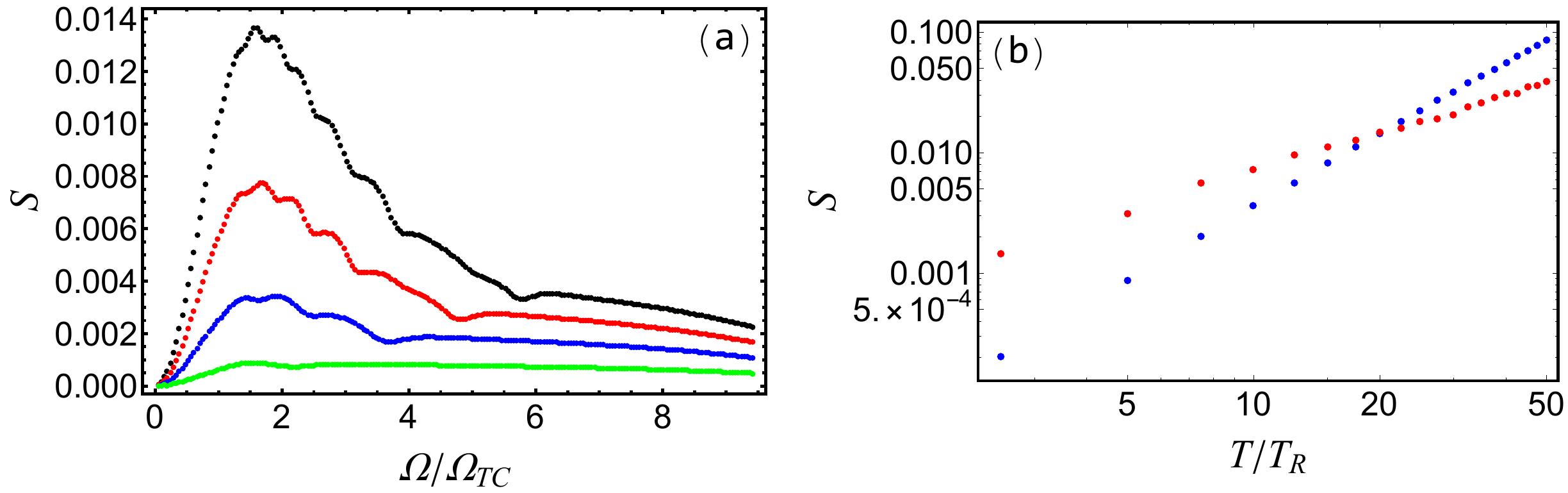}
\caption{(a) The dependence of the sensitivity $S$ on the coupling strength calculated over the observation time $T=5T_R$ (green dots), $T=10T_R$ (blue dots), $T=15T_R$ (red dots) and $T=20T_R$ (black dots). (b) The dependence of sensitivity on time in double logarithmic scale at coupling strengths $\Omega = 0.5 \Omega_{TC}$ (blue dots) and $\Omega = 7\Omega_{TC}$ (red dots). The magnitude of the perturbation is $\varepsilon = 10^{-5}$. $N=50$, $\delta\omega=4\cdot10^{-3}\omega_0$, $g=6\cdot10^{-3}\omega_0$.}
\label{fig2}
\end{figure*}

\begin{figure}[htbp]
\centering\includegraphics[width=1.0\linewidth]{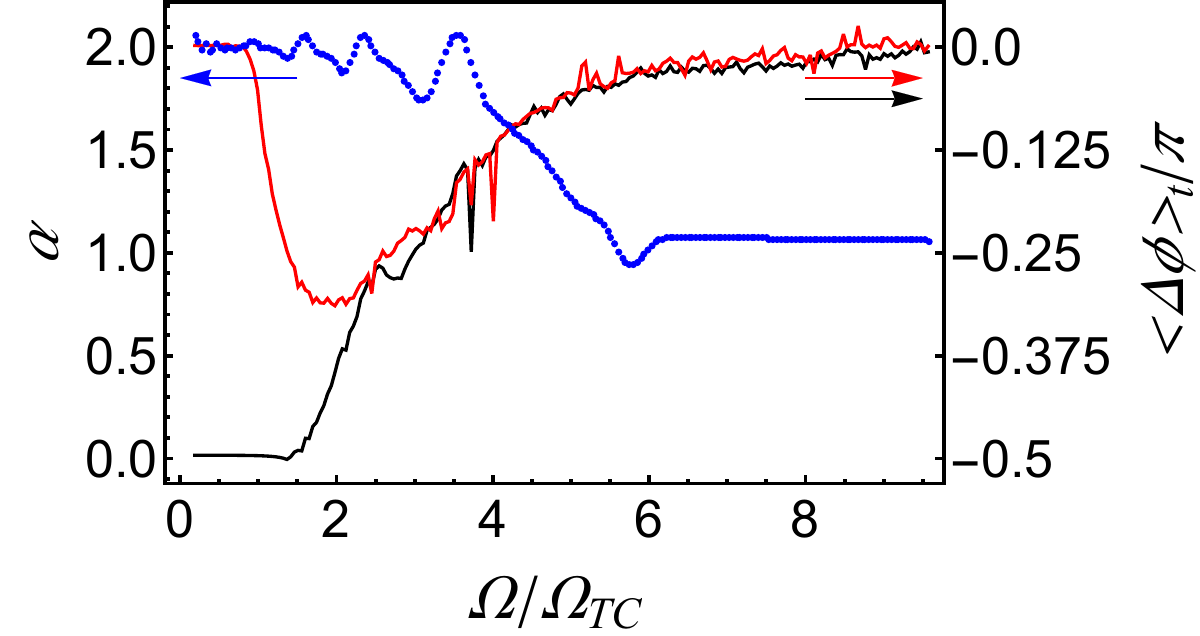}
\caption{The dependence of the observation time exponent $\alpha$ of the sensitivity $S \thicksim T^{\alpha}$ on the coupling strength (blue dots). The dependence of phase difference between the atom's and cavity's amplitudes on coupling strength averaged over the time of one bypass (black line) and two bypasses (red line). The magnitude of the perturbation is $\varepsilon = 10^{-5}$. $N=50$, $\delta\omega=4\cdot10^{-3}\omega_0$, $g=6\cdot10^{-3}\omega_0$.}
\label{fig3}
\end{figure}

Figure~\ref{fig2} (a) demonstrates the dependence of the value $S$ on the coupling strength $\Omega$ at fixed magnitude of the perturbation $\varepsilon$ for different observation times $T=mT_{R}=m\frac{2\pi}{\delta\omega}$, where $m=5, 10, 15, 20$. It is seen that at $\Omega \sim \Omega_{TC}$ the sensitivity to the perturbation, $S$, increases noticeably more with increasing observation time than in the area $\Omega >> \Omega_{TC}$. Our calculations show that the sensitivity $S$ increases in the power law $S \sim T^\alpha$ with increasing observation time $T$ [Figure~\ref{fig2}(b)]. Figure~\ref{fig3} shows the dependence of the observation time power $\alpha$ on the coupling strength at a fixed magnitude of the perturbation $\varepsilon$ (blue dots). It is seen that when $\Omega < 4  \Omega_{TC}$, an approximately quadratic dependence of the sensitivity on the observation time takes place ($\alpha \sim 2$). When $\Omega > 6 \Omega_{TC}$, the dependence of the sensitivity on the observation time is linear ($\alpha \sim 1$). The type of power law depends on the regime (the time crystal regime, etc) in which the system is.

To determine the regime taking place at a given coupling strength, following \cite{15} we calculate the phase difference between the atom's and single-mode resonator amplitudes. When $\Omega < \Omega_{TC}$, the atom's amplitude oscillates with a period of two bypasses $2T_R$ and there is non-zero phase difference between the atom's and single-mode resonator amplitudes averaged over a time of one bypass $T_R$ (the black solid line in Figure~\ref{fig3}). The non-zero phase difference value indicates that the system does not return to its initial state within a time of one bypass, $T_R$. However, when averaging is carried out over a time of two bypasses $2T_R$, the average phase difference is zero (the red solid line in Figure~\ref{fig3}). That is, the system returns to its initial state within a time of two bypasses. This behavior corresponds to the time crystal regime. At large coupling strengths $\Omega >> \Omega_{TC}$ the system oscillates with a period of one bypass and the time crystalline order is completely disrupted. In this regime, the average phase difference is zero in both cases [Figure~\ref{fig3}]. That is, the system manages to return to its initial state during a time of one bypass. This behavior corresponds to the normal state (the state with unbroken time translation symmetry). In addition, there is a transition area between these two cases, which is characterized by complex dynamics and where is no time crystalline order, but its residual traces. 

Thus, our calculations show that for the coupling strengths corresponding to the time crystal regime and the transition area, the dependence of sensitivity $S$ on the observation time is quadratic ($\alpha \sim 2$) [Figure~\ref{fig3}]. Outside these areas, the dependence on the observation time becomes linear ($\alpha \sim 1$) [Figure~\ref{fig3}]. Our calculations demonstrate that the sensitivity in the area $\Omega \sim \Omega_{TC}$ becomes greater than in the area $\Omega >> \Omega_{TC}$, i.e., the quadratic time dependence exceeds the linear one, when the observation time is of order or greater than $10 T_R$ (Figure~\ref{fig2} (a)). However, it is important to note here that since the probability amplitude of the atom is always limited to one, the quadratic scaling of sensitivity on the observation time works only for limited time intervals (see Figure~\ref{fig1a} in Appendix B). Then saturation sets in and the gain in sensitivity disappears.

\section{Influence of memory on the dependence of sensitivity on observation time}

In the time crystal regime, a quadratic dependence of the perturbation magnitude sensitivity on the observation time is observed. This is due to the fact that in the time crystal regime the atom's state retains memory of its initial state. Indeed, as can be seen in Figure~\ref{fig1}, in the time crystal regime, the state of the atom does not have sufficient time to completely dissipate. To understand how maintaining an atom's state affects sensitivity, consider the interaction of the atom with the electromagnetic pulses it emits into multimode resonator. The radiation of the atom leads to the excitation of the electromagnetic pulses that propagate in clockwise and counterclockwise directions in the ring resonator. After a single pass through the ring resonator, the difference between the electromagnetic pulses emitted by the atom and propagating in opposite directions is equal to $\Delta \varphi  = {\varphi _ + } - {\varphi _ - } \sim {\Omega _{rot}}{T_R}$. Returning to the atom, the electromagnetic states act on the state of the atom. This action depends on the phase difference between the pulses. As a result, the state of the atom in the unperturbed case will differ from the state in the perturbed one by an amount that depends on ${\Omega _{rot}}{T_R}$ (i.e. $S\left( {T={T_R}} \right)$is a function of ${\Omega _{rot}}{T_R}$). In the case where the difference between the perturbed and unperturbed cases is small, the linear approximation is applicable $S\left( {{T_R}} \right) = A\,{\Omega _{rot}}{T_R}$, where $A$ is the proportionality coefficient. After the second pass through the resonator, the phase difference between the pulses becomes equal to $\Delta \varphi  \sim 2{\Omega _{rot}}{T_R}$. In this case, the action of the pulses on the atom leads to an additional difference between the perturbed and unperturbed systems proportional to $2{\Omega _{rot}}{T_R}$. If the state of the atom does not have time to decay during the observation time, then the perturbations are summed up and the dependence of the sensitivity on time is determined by the expression
$S\left( T \right) = A\,\int\limits_0^T {{\Omega _{rot}}{T_R}d{T_R}}  \sim {T^2}$. Such a dependence takes place in the time crystal regime.
If the state of the atom decays during the the time of one bypass of the ring resonator, then the difference in the states of the atom for the perturbed and unperturbed systems is determined only by the last action of the electromagnetic pulses. In this case, $S\left( T \right) \sim \Delta\varphi \sim T$. Such behavior is observed outside the time crystal regime.

Thus, in the time crystal regime, the current state of the atom is determined by the integral over all previous states. And therefore, the sensitivity is determined by contributions over the entire observation time ($S\left( T \right) \sim T^2$). On the contrary, outside the time crystal regime, the state of the atom completely dissipates during the one bypass of the ring resonator and so the sensitivity is determined by contributions only for the finite time interval ($S\left( T \right) \sim T$). In the transition area between these regime, the sensitivity depends quadratically on the observation time. This indicates that in the transition area the system partially retains the memory of the initial state of the atom and the current state of the atom is determined by the integral over previous states. This is due to the complex dynamics of the system in the transition area, in which $\vert C_{\sigma} \vert^{2}$ remains significantly different from zero at all times [Figure~\ref{fig1}(c)].

\begin{figure}[htbp]
\centering\includegraphics[width=1.0\linewidth]{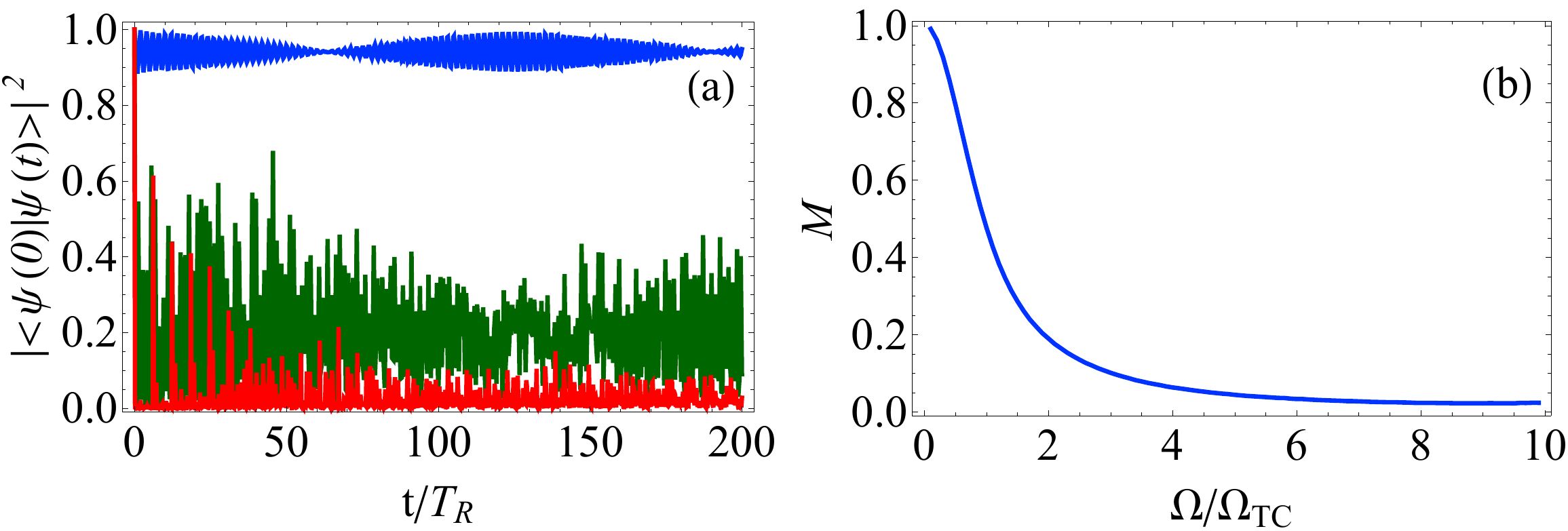}
\caption{(a) The dependence of $p\left( t \right) = {\left| {\left\langle {{\Psi \left( 0 \right)}} \mathrel{\left | {\vphantom {{\Psi \left( 0 \right)} {\Psi \left( t \right)}}}  \right. \kern-\nulldelimiterspace}  {{\Psi \left( t \right)}} \right\rangle } \right|^2}$ on time when the system is in the time crystal regime $\Omega = 0.25 \cdot \Omega_{TC}$ (the blue line), in the transition area $\Omega = 2 \cdot \Omega_{TC}$ (the green line) and outside the time crystal regime $\Omega = 8 \cdot \Omega_{TC}$ (the red line). (b) The dependence of $M$ (see the Eq.~(\ref{eq:8})) on $\Omega$. All other parameters as in Figure~\ref{fig1}.}
\label{fig5}
\end{figure}

To quantitatively describe the memory retention in the system, we consider the quantity $p\left( t \right) = {\left| {\left\langle {{\Psi \left( 0 \right)}} \mathrel{\left | {\vphantom {{\Psi \left( 0 \right)} {\Psi \left( t \right)}}}  \right. \kern-\nulldelimiterspace}  {{\Psi \left( t \right)}} \right\rangle } \right|^2}$. This quantity determines the probability that at moment of time $t$ the system is in the initial state.
Our calculations show that in the time crystal regime $p\left( t \right)$ takes values close to $1$ at all times [Figure~\ref{fig5}(a)]. With an increase in the coupling strength, the value of $p\left( t \right)$ decreases [Figure~\ref{fig5}(a)]. For a more accurate memory characteristic, we calculate the average value of $p\left( t \right)$ over the time interval from $t_1$ to $t_2$:

\begin{equation}
M = \int_{t_1}^{t_2} p\left( t \right) \,dt
\label{eq:8}
\end{equation}
where $t_{1,2} > > T_R$. Numerical simulations show that $M$ decreases monotonically with increasing $\Omega$ [Figure~\ref{fig5}(b)]. This quantity characterizes the presence of memory of the initial state in the system. It is seen that memory retention takes place in the time crystal regime and in the transition region [Figure~\ref{fig5}(b)]. This result is consistent with the sensitivity behavior [Figures~\ref{fig2},\ref{fig3}].

Our calculations show that for $T> 5 T_R$ the sensitivity has a maximum when $\Omega \sim 2 \Omega_{TC}$ [Figure~\ref{fig2}(a)]. The maximum sensitivity is associated with two effects. Firstly, since only the modes of ring resonator are perturbed, the effect of perturbations on the atomic dynamics is directly determined by the coupling strength between the atom and the rest of system that is proportional to $\Omega$. The greater the coupling strength, the greater the influence of the perturbations, which determines the sensitivity. Secondly, the time crystalline order and the memory retention are present at low coupling strength $\Omega \sim \Omega_{TC}$. An increase in the coupling strength leads to the destruction of the time crystalline order and the memory retention [Figure~\ref{fig5}]. Preserving the memory of the initial state enhances the sensitivity. Thus, the sensitivity is affected by two effects. As the coupling strength increases, one of them decreases the sensitivity and the other increases. Accordingly, there is a maximum sensitivity when one effect begins to exceed the other. 

Note that the sensitivity, $S$, is determined by the ratio of time-averaged $\vert C_{\sigma,\varepsilon \neq 0} \vert^{2}$ and $\vert C_{\sigma,\varepsilon = 0} \vert^{2}$. Similar dependencies on the observation time, $T$, are obtained for a quantity determined by the ratio of non-averaged $\vert C_{\sigma,\varepsilon \neq 0} \vert^{2}$ and $\vert C_{\sigma,\varepsilon = 0} \vert^{2}$  ($1- \vert C_{\sigma,\varepsilon \neq 0} (t=T) \vert^{2} / \vert C_{\sigma,\varepsilon= 0} (t=T) \vert^{2}$). However, the obtained dependencies on the observation time are not universal and may be different for other quantities characterizing the sensitivity of the system to perturbations.

\section{Conclusion}
In conclusion, we consider the composite atom-cavity system interacting with the ring resonator. In this structure there is a parameter range, in which the spontaneous breaking of time translation symmetry is possible. We demonstrate that the time crystal regime allows a quadratic increase in sensitivity with increasing observation time. This is due to the fact that in the time crystal regime the system is able to retain the memory of its initial atom's state. At values of coupling strength greater than the critical coupling strength, when the time crystal regime is disrupted, the atom's state begins to destroy significantly, losing memory of its initial state. Eventually, the dependence of sensitivity on the observation time becomes linear. The quadratic increase in sensitivity with increasing observation time opens up a new ways to create measuring devices, for example, such as optical gyroscopes.

\section*{Acknowledgments}
The study was financially supported by a Grant from Russian Science Foundation (Project No. 23-42-10010, https://rscf.ru/en/project/23-42-10010/). E.S.A thanks foundation for the advancement of theoretical physics and mathematics “Basis”.

\section*{Appendix A. The system with a large number of the atoms}
We consider the system with $m$ atoms. We use the following Hamiltonian for the analysis \cite{23}:

\begin{equation}
\begin{array}{l}
\hat H = \sum\limits_{k = 1}^{m} {\omega _0}\hat \sigma_k^\dag {{\hat \sigma_k}} + {\omega _0}\hat a^\dag {{\hat a}} + \sum\limits_{k = 1}^{m} \Omega_0 (\hat a {{\hat \sigma_k^{\dag} }} + \hat a^{\dag} {{\hat \sigma_k}}) + \\
\sum\limits_{j = j_0-N/2}^{j_0+N/2} {{\omega_j^{+}}\hat \alpha_j^\dag {{\hat \alpha}_j}} + \sum\limits_{j = j_0-N/2}^{j_0+N/2} {{g_j}(\hat a {{\hat \alpha}_j^{\dag}} + \hat a^{\dag} {{\hat \alpha}_j})}  + \\
\sum\limits_{j = j_0-N/2}^{j_0+N/2} {{\omega_j^{-}}\hat \beta_j^\dag {{\hat \beta}_j}} +\sum\limits_{j = j_0-N/2}^{j_0+N/2} {{g_j}(\hat a {{\hat \beta}_j^{\dag}} + \hat a^{\dag} {{\hat \beta}_j})}
\end{array}
\label{eq:1AA}
\end{equation}
where $\hat \sigma_k$ and $\hat\sigma_k^{\dag}$ are annihilation and creation operators of $k$-th two-level atom that obey the fermionic commutation relation $\{\hat\sigma_k,\hat\sigma_k'^{\dag}\} = \delta_{k,k'}$. $\Omega_0$ is the coupling strength between the single-mode cavity and each of the atom. The remaining quantities are the same as in Hamiltonian (\ref{eq:1}). The system dynamics are determined by the time-dependent Schr\"{o}dinger equation for a wave function $\vert\Psi (t)\rangle$. We assume that there is only one excitation quantum in the system. We look for the wave function in the following form:

\begin{equation}
\begin{array}{l}
\vert \Psi(t)\rangle = \sum\limits_{k = 1}^{m} C_{k}(t)\vert e_k,0,0,0\rangle + C_{a}(t)\vert g,1,0,0\rangle + \\
\sum\limits_{j = j_0-N/2}^{j_0+N/2} {C_{j}^{+}(t)\vert g,0,1_{j}^{+},0\rangle} + \sum\limits_{j = j_0-N/2}^{j_0+N/2} {C_{j}^{-}(t)\vert g,0,0,1_{j}^{-}\rangle} 
\end{array}
\label{eq:2AA}
\end{equation}
where $\vert e_k,0,0,0 \rangle$, $\vert g,1,0,0 \rangle$, $\vert g,0,1_{j}^{+},0\rangle$ and $\vert g,0,0,1_{j}^{-}\rangle$ are the states, in which the excitation quantum is in one of the atoms, in the single-mode cavity, in one of the ring resonator "clockwise" and "counterclockwise" modes, respectively. $C_{k}(t)$, $C_{a}(t)$, $C_{j}^{+}(t)$ and $C_{j}^{-}(t)$ are the amplitudes of probability of finding the excitation quantum in one of the atoms, in the single-mode cavity, or in one of the ring resonator "clockwise" and "counterclockwise" modes, respectively.

We can obtain the closed system of equations for the probability amplitudes after substituting the wave function~(\ref{eq:2}) into the Schr\"{o}dinger equation with Hamiltonian~(\ref{eq:1}):

\begin{equation}
\frac{{d{C_{k}}}}{{dt}} =  - i{\omega _0}{C_{k}} - i\,\Omega_0 {C_{a}} 
\label{eq:3AA}
\end{equation}

\begin{equation}
\begin{array}{l}
\frac{{d{C_a}}}{{dt}} =  - i{\omega _0}{C_a} - i \sum\limits_{k = 1}^{m} \,\Omega_0 {C_{k}} \\
-i\sum\limits_{j = j_0-N/2}^{j_0+N/2} {g{C_j^{+}}} -i\sum\limits_{j = j_0-N/2}^{j_0+N/2} {g{C_j^{-}}}
\end{array}
\label{eq:4AA}
\end{equation}

\begin{equation}
\frac{{d{C_j^{+}}}}{{dt}} =  - i{\omega_j^{+}}{C_j^{+}} - ig{C_a}
\label{eq:5AA}
\end{equation}

\begin{equation}
\frac{{d{C_j^{-}}}}{{dt}} =  - i{\omega_j^{-}}{C_j^{-}} - ig{C_a}
\label{eq:6AA}
\end{equation}
By introducing a quantity $C_\sigma = \frac{1}{\sqrt{m}} \sum\limits_{k = 1}^{m} {C_{k}}$ and a coupling strength $\Omega = \sqrt{m} \Omega_0$, we get the equations that coincide with equations~(\ref{eq:3})-(\ref{eq:6}). Considering that $C_{\sigma}(0)=1$ and $C_a (0)=C_j^{+}(0)=C_j^{-}(0)=0, j=j_0-N/2,...,j_0+N/2$, we obtain a case equivalent to the one considered in the main part of the article.

\section*{Appendix B. Long-term evolution properties and discussion of experimentally realized parameters}
\begin{figure}[htbp]
\centering\includegraphics[width=0.8\linewidth]{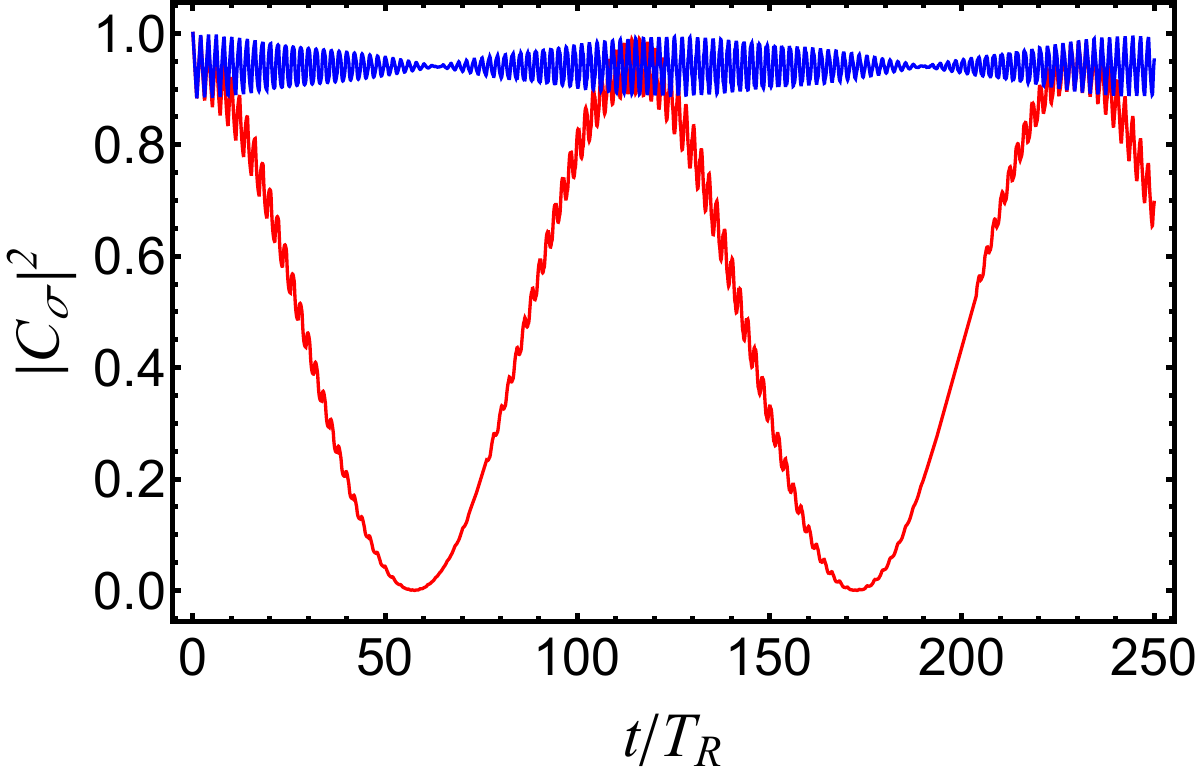}
\caption{Dependence of the atom's probability density $\vert C_{\sigma}(t)\vert^{2}$ on time in the case of zero perturbations $\varepsilon = 0$ (blue solid line) and in the case of non-zero perturbation $\varepsilon = 10^{-4}$ (red solid line) in the time crystal regime $\Omega = 0.25 \cdot \Omega_{TC}$. We consider $N=50$, $\delta\omega = 4\cdot 10^{-3} \omega_0$, $g=6\cdot 10^{-3}\omega_0$.}
\label{fig1a}
\end{figure}

It is important to note that time crystalline order is stable at long-term evolution. Figure~\ref{fig1a} shows that oscillations with the period of two bypasses of the ring resonator ($2T_R$) are persistent at times much greater than the time of one bypass. The dynamics of the perturbed system (red line in Figure~\ref{fig1a}) also demonstrate the persistent oscillations at long-term evolution. The perturbations in modes' frequencies lead to appearance of additional oscillations with period much greater than the time of one bypass. This period is inversely proportional to the magnitude of perturbation $T \sim\varepsilon^{-1}$. And this period also limits the area of the quadratic scaling of the sensitivity.

In the ring resonator, the frequency shifts of "clockwise" and "counterclockwise" modes ($\omega _j^ \pm  =  j\,\delta \omega \left( {1 \pm \varepsilon } \right)$) occurs due to rotation of the system with angular frequency ${\Omega _{rot}}$. In this case, $\varepsilon$ is determined by the expression $\varepsilon  = {\Omega _{rot}} \cdot R/c$ (see discussion in Section Model), where $c$ is the speed of light and $R$ is the radius of the ring resonator, which is expressed through the parameters used in the calculations as $R=c/ \delta \omega$. Thus, $\varepsilon  = {\Omega _{rot}} / \delta \omega$. When $\Omega _{rot} = 1 s^{-1}$ and the transition frequency of atoms lies in the optical range ($\omega_0 \approx 3 \cdot 10^{15} s^{-1}$; $\delta \omega = 4\cdot 10^{-3} \omega_0$; $R \sim 10^{-4} m$), $\varepsilon \sim 10^{-13}$. Our calculations show that the quadratic scaling of the sensitivity holds for all values of $\varepsilon < 10^{-3}$. With realistic parameters, the period, $T$, limited the area of the quadratic scaling of the sensitivity, is much greater than $T_R$. Thus, the predicted dependencies of sensitivity on observation time can be observed in real systems.

\nocite{*}

\bibliography{apssamp}

\end{document}